\documentstyle[12pt]{article}
\begin{document}

\sloppy

\title{Length and time scale divergences at the 
magnetization-reversal transition in the Ising model}

\author{R. B. Stinchcombe$^{1,2},$ A. Misra$^2$ and B K Chakrabarti$^2$\\
$^1$ Department of Physics$,$ University of Oxford, \\
1 Keble Road$,$ Oxford$,$ OX1 3NP$,$ UK \\
$^2$ Saha Institute of Nuclear Physics, \\ 
${\rm 1/AF}$ Bidhan Nagar$,$ Calcutta 700064$,$ India.}

\maketitle

\begin{abstract}
The divergences of both the length and time scales, at the magnetization-
reversal transition in Ising model under a pulsed field, have been studied 
in the linearized limit of the mean field theory. Both length and time
scales are shown to diverge at the transition point and it has been 
checked that the nature of the time scale divergence agrees well with the 
result obtained from the numerical solution of the mean field equation of motion.
Similar growths in length and time scales are also observed, as one approaches the 
transition point, using Monte Carlo simulations. However,
these are not of the same nature as the mean field case.
Nucleation theory provides a qualitative argument which explains the nature
of the time scale growth. To study the nature of growth of the 
characteristic length scale, we have looked at the cluster size distribution
of the reversed spin domains and defined a pseudo-correlation length which 
has been observed to grow at the phase boundary of the transition.
\end{abstract}

\newpage

The dynamic response of pure Ising systems to time dependent magnetic 
fields is currently being studied intensively \cite{rmp}.
In particular, the response of Ising systems to pulsed fields has recently 
been investigated \cite{pos,pep1,pep2}. The pulse can be either 
``positive'' or ``negative''. At temperatures $T$
below the critical temperature $T_c$ of the 
corresponding static case (without any external field), the  majority of the 
spins orient themselves
along  a particular direction giving rise to the prevalent order. 
In the following, we denote by positive (or negative) pulse an
 external field pulse  applied along (or opposite)  the direction of the 
existing order. 
 The effects
 of a positive pulse can be  
analyzed by extending appropriately the finite size scaling technique to
this finite time window case \cite{pos}, and it does not involve
any new transition or introduce any new thermodynamic
scale into the problem. The negative field pulse, on the other hand,
induces a new dynamic 
``magnetization-reversal'' transition, involving completely new length
and time scales \cite{pep1,pep2}. In fact, we believe, the spontaneously occurring
dynamic symmetry-breaking transition in Ising models under (high
frequency) external oscillating fields \cite{rmp,rik} occurs actually during this
``negative'' pulse period (and not during the ``positive'' pulse period;
compared to the  instantaneous existing order in the system), and the
universality classes of these two transitions are identical. 
We report here the results of an investigation on the nature of the
characteristic length and time scales involved in  
this dynamic magnetization-reversal transition in an Ising model
under the negative pulsed field.

In the absence of any symmetry breaking field, for temperatures below the critical 
temperature of the corresponding
static case ($T < T_c$), there are two equivalent free energy minima with
 average magnetizations 
$+m_0$ and $-m_0$. If in the ordered state the equilibrium magnetization is 
$+m_0$ (say) and a very weak pulse is applied in the direction 
opposite to the existing order, then temporarily during the pulse
period the free energy minimum with magnetization $-m_0$ will be brought down 
compared to that with $+m_0$. If this asymmetry is made permanent, then 
any non-vanishing field (strength), which is 
responsible for the asymmetry,  would eventually induce 
a transition from $+m_0$ to $-m_0$ (in the limit of vanishing
field strength). Instead, if the field is applied 
in the form of a pulse, the asymmetry 
in the free energy wells is removed after a finite
period of time. In that case, the point of interest lies in the combination of 
the pulse height or strength ($h_p$) 
and its width or duration ($\Delta t$) that can 
give rise to the transition from $+m_0$ to $-m_0$. We call this a 
magnetization-reversal transition. A crucial point about the 
transition is that it is not necessary that the system should attain its final 
equilibrium
magnetization $-m_0$ during the presence of the pulse; the combination of $h_p$
and $\Delta t$ should be such 
that the final equilibrium state is attained at any
subsequent time, even a long time after the 
pulse is withdrawn (see Fig. 1) . The ``phase boundary'',
giving the minimal combination of $h_p$ 
and $\Delta t$ necessary for the transition,
depends on the temperature. As $T \rightarrow T_c$, the magnetization reversal 
transition
occurs at lower values of $h_p$ 
and/or $\Delta t$ and the transition disappears at
$T \ge T_c$.

In the present paper we present an argument that this 
dynamic transition corresponds to infinite time  and
length scales, all along the phase boundary in the $h_p-\Delta t$ 
plane at any temperature at T $\ < T_c$. We show that the relaxation 
time $\tau$ and the correlation length $\xi$ both diverge as one 
approaches the phase boundary. In the mean field case, we show (using 
equations of motion linearized in the magnetization) that
$$\tau \sim \ln \left( \frac{1}{m_w} \right)~~~~~~ {\rm and} ~~~~~~~~~
\xi \sim \sqrt{\ln \left( \frac{1}{m_w}\right)}$$
where $m_w$ is the ``order-parameter'' for the transition,
given by the magnetization at the time of withdrawal of
the pulse, starting from $m_0$, the equilibrium  magnetization at the
temperature $T (< T_c)$ (see Fig. 1). It may be noted that
$m_w(T, h_p, \Delta t) = 0$ at the phase boundary of the magnetization-reversal
transition.

We also show that $\xi$ and 
$\tau$ grow sharply as one approaches the phase boundary in the Monte Carlo case as
well, although the nature of the growths are different from the mean field case. 
We also study the shapes and sizes of the reversed spin domains 
as one approaches the spin-reversal transition phase 
boundary in the Monte Carlo case. We compare the observed growth in the 
relaxation time in this case with that predicted by the nucleation theory. 


The Ising model in the presence of an external magnetic field is described
by the Hamiltonian

\begin{equation}
H=- \frac{1}{2} \sum _{(ij)}J_{ij}S_{i}S_{j}-\sum _{i}h_iS_{i}, 
\end{equation}
where \( S_{i} \) denotes the spin at \( i \)th site,
\( J_{ij} \) is the cooperative interaction between the spins at sites
$i$ and $j$ and $(...)$ denotes the nearest-neighbour
pairs. Here $h_i$ is the external field, allowed to be time
dependent, and also site-dependent to allow investigation of 
separation-dependent correlations.
The free energy of the system in the Bragg-Williams approximation is
given by\cite{brout}

\begin{equation}
F=-\frac{1}{2}\sum_{(ij)} J_{ij} m_im_j -\sum_i h_i m_i
+\sum_i \frac{T}{2}[\ln (1-m_i^{2})+
m_i\ln \left( \frac{1+m_i}{1-m_i}\right)- 2\ln 2], 
\end{equation}
with $m_i = \langle S_i \rangle$, where $\langle ... \rangle$ denotes the
thermal average.
In the presence of a time and site-dependent field, the time dependent magnetization
satisfies the Langevin equation

\begin{equation}
\frac{dm_i}{dt} = - \frac {\lambda}{T} \frac{\delta F} {\delta m_i} =
\lambda \left[ \sum_j K_{ij} m_j(t) + \frac {h_i(T)}{T} - \frac {1}{2}
\ln \left(\frac {1+m_i(t)}{1-m_i(t)} \right)\right], 
\end{equation}
where $K_{ij} = J_{ij}/T$ and $\lambda$ is a constant. Differentiation with
the space and time dependent magnetic field $h_i(t)$ generates the space and
time dependent susceptibility. After the differentiation we can set $h_i(t) = h(t)$
to obtain results for a pulsed field uniform in space. Then $m_i(t) \rightarrow m(t)$ gives

\begin{equation}
\frac{dm(t)}{dt}=\lambda \left[ K(0)m(t)+\frac{h(t)}{T}-\frac{1}{2}\ln
 \left( \frac{1+m(t)}{1-m(t)}\right) \right]. 
\label{eq:d}
\end{equation}
The resulting equation for the susceptibility, in the Fourier space, is 

\begin{equation}
\frac{d\chi _{q}(t)}{dt}=\lambda \left[ K(q)-\frac{1}{1-m^{2}(t)}\right] 
\chi _{q}(t)+\frac{\lambda}{T} \delta (t-t'). 
\label{eq:e}
\end{equation}
Here, $K(q)$ is the Fourier transform of $K_{ij}$; for small $q$,
 \( K(q) \simeq K(0)(1-q^{2}) \); in the mean field theory $K(0) = T_c/T$.
Using (\ref{eq:d}) and (\ref{eq:e}), we can write
\begin{equation}
\frac{d\chi _{q}(t)}{dm(t)}=\frac{\left[ K(q)-\frac{1}{1-m^{2}(t)}\right] 
\chi _{q}(t)}{K(0)m(t)-\frac{h_{p}}{T}-\frac{1}{2}\ln \left[ \frac{1+m(t)}
{1-m(t)}\right] }.
\end{equation}
In the limit when \( m(t) \) is small, retaining up to the linear term in 
$m(t)$,
\begin{equation}
\frac{d\chi _{q}(t)}{dm(t)}=\frac{[K(q)-1]\chi _{q}(t)}
{[K(0)-1]m(t)-\frac{h_{p}}{T}}. 
\label{eq:g}
\end{equation}

This equation can now be solved in the three different time zones (Fig. 1): namely, in
the equilibrium regime before the application of the pulse
where $m = m_0$ (regime I), the 
(nonequilibrium) pulsed period regime, at the end of 
which $m = m_w$ (regime II), and the regime after the
pulse is withdrawn (regime III) when the system eventually returns to
equilibrium (with $m(t \rightarrow \infty) = -m_0$ if the transition occurs,
or $= m_0$ if it does not). Hence in regime II and III, we get the
non-equilibrium susceptibility $\chi_q$ as a function of $m(t)$. The
solution of (\ref{eq:d}) also gives the non-equilibrium magnetization $m(t)$, and hence
we can also arrive at $\chi_q (t)$. Noting that
$ \chi _{q}(t)=\chi _{q}^{s} $ when \( m(t)=m_{0} \), at the start of regime II,
 where \( m_{0} \) and \( \chi ^{s}_{q} \) are equilibrium
values of the magnetization and susceptibility respectively, we can
integrate (\ref{eq:g}) in that regime to obtain

\begin{equation}
\frac{\chi _{q}(t)}{\chi _{q}^{s}}=\left[ \frac{m(t)-\Gamma }
{m_{0}-\Gamma }\right] ^{a_{q}}, 
\label{eq:h}
\end{equation}
where 
$$\Gamma =\frac{h_{p}/T}{K(0)-1} \eqno (8a)$$
and
$$a_{q}=\frac{K(q)-1}{K(0)-1}. \eqno (8b)$$
Also integrating the linearized version of (\ref{eq:d})  in region II, one gets

\begin{equation}
m(t)=\Gamma +(m_{0}-\Gamma )\exp [\lambda b(t-t_{0})], 
\end{equation}
where \( b=K(0)-1 \). At the end of region II, the value of magnetization
is given by

\begin{equation}
m_w = m(t_0+\Delta t) = \Gamma + (m_0 - \Gamma) \exp(\lambda b \Delta t).
\end{equation}
The eqn. (\ref{eq:h}) can therefore be written as

\begin{equation}
\frac{\chi _{q}(t)}{\chi _{q}^{s}}=\exp 
(\lambda ba_{q}t)=\exp [(K(q)-1)\lambda t].
\end{equation}

In regime III, however, \( h(t)=0 \) and the 
(initial) boundary condition is \( m(t_{0}+\Delta t)=m_{w} \). Integrating
(\ref{eq:g}) in this regime one gets
\[
\frac{\chi _{q}(t)}{\chi _{q}(t_{0}+\Delta t)}=\left[ \frac{m(t)}{m_{w}}\right] ^{a_{q}}\]
or

\begin{equation}
\chi _{q}(t)=\chi _{q}^{s}\exp [\lambda (K(q)-1)(t_{}+\Delta t)]\left[ 
\frac{m(t)}{m_{w}}\right] ^{a_{q}}, 
\end{equation}
where use has been made of the eqn. (\ref{eq:h}). Concentrating on the dominating 
\( q \)-dependence of the susceptibility, one can write
\begin{equation}
\chi _{q}(t)\sim \chi _{q}^{s}\exp [-q^{2}\xi ^{2}], 
\end{equation}
where the correlation length \( \xi  \) is defined as
\begin{equation}
\xi \equiv \xi (m_{w})=\left[ \frac{\ln (1/m_{w})}{1-T/T_{c}}\right] 
^{\frac{1}{2}}. 
\label{eq:m}
\end{equation}
This is one of the principal results of this paper, and it shows that the 
characteristic length $\xi$ diverges as the order parameter $m_w$ goes to
zero.

Consider now the \( t \) dependence arising 
in \( \chi _{q=0}(t) \) through the factor \( m(t)^{a_{q}} \). Solving
(\ref{eq:d}) in regime III yields
\begin{equation}
m(t)=m_{w}\exp [\lambda b\{t-(t_{0}+\Delta t)\}], 
\label{eq:n}
\end{equation}
 which shows that long time is required to
attain moderate values of \( m(t) \) starting from low values of \( m_{w} \). Especially,
starting from time \( t=t_{0}+\Delta t \), the time taken by the system to reach the final
equilibrium value is defined as the relaxation time \( \tau  \) of the system.
Therefore from (\ref{eq:n}) we can write

\begin{equation}
\tau =\frac{1}{\lambda }\left( \frac{T}{T_{c}-T}\right) 
\ln \left( \frac{m_{0}}{m_{w}}\right). 
\label{eq:o}
\end{equation}
 The growth of the time scale occurs in \( \chi _{q=0}(t) \) too 
through the \( m(t) \) dependence
: 
$$\chi _{q=0}(t)\sim \left[ \frac{m(t)}{m_{w}}\right] 
^{a_{q=0}}\sim \exp [\lambda b\{t-(t_{0}+\Delta t)\}].$$
Eqn. (\ref{eq:m}) and (\ref{eq:o}) can be used to establish a relationship between
\( \tau  \) and \( \xi  \) :
\begin{equation}
\tau \sim \ln \left( \frac{1}{m_{w}}\right) \sim 
\frac{T}{T_{c}}\xi ^{2}. 
\label{eq:p}
\end{equation}
This corresponds to critical slowing, with the characteristic time
diverging with the characteristic length with the dynamical critical exponent 
\[
z=2.\]

The above results are obtained in the linearized limit of the mean field eqns. of
motion (\ref{eq:d}) and (\ref{eq:e}). We also measured, solving the full dynamical equation (\ref{eq:d})  numerically,
 the relaxation time $\tau$ by
computing the time required by $m(t)$ to reach the final equilibrium value $\pm m_0$,
with an accuracy of $O(10^{-4})$, from the time of withdrawal of the pulse (in regime
III). Fig. 2 shows that this $\tau$ indeed diverges as one approaches the phase boundary,
where $m_w = 0$. In fact, the numerical results are observed to fit
very well with the  analytic result (\ref{eq:p}) (shown by the solid line
in Fig. 2). 


The divergence of both the time and length scale were also investigated 
at low temperatures 
by employing Monte Carlo methods. Simulations \cite{pep2} on a square lattice
of typical size $L=200$ with periodic boundary conditions indicated an
exponential growth of the time scale : 
\begin{equation}
\tau \sim \exp[-c(T) \mid m_w \mid], 
\label{eq:q}
\end{equation}
where $c(T)$ is a constant depending on temperature only. 
Further, finite size scaling of the order parameter relationship 
\begin{equation}
m_w \sim~ \mid h_p - h_p^c \mid^\beta 
\end{equation}
is consistent\cite{pep2} with $\beta = 0.90
\pm 0.02 $ and with a correlation length divergence with $\nu = 1.5 \pm 
0.3$. (Here $h_p^c$ is the critical value of the pulse field $h_p$, making 
$m_w=0$ at the end of regime II). These results qualitatively compare with the 
divergence of scales at the transition point predicted by the mean field treatment. 
However, the growths of the time 
and length scales are quantitatively of different nature to that
of the mean field case, because at low temperatures droplet growth is a 
dominant mechanism. The growth of droplets of size $l$ is 
associated with an activation energy\cite{pep2} $E(l) = -2h_pl^d + \sigma l^{d-1}$,
where $\sigma$ is the surface tension. Using the relationship between
$l$ and $h_p$ at the energy minimum together with (18) at small $m_w$ gives
a characteristic time
\begin{equation}
\tau ~~\sim ~~\exp\left[\frac{1}{T} h_p^{1-d}\right] 
 ~~\sim ~~\exp[-c_1(T) \mid m_w \mid^{1/\beta} (h_p^c)^{d-2}]. 
\end{equation}
Since $\beta$ is close to unity, this is consistent with the observed 
relation (\ref{eq:q}). 

The typical size of cluster or domain of reversed spins provides a qualitative idea about the
correlation length of the system. In order to study the growth of 
the typical reversed-spin domain size, we define a pseudo-correlation length 
$\tilde \xi $ as follows : 
\begin{equation}
\tilde \xi^2 = \frac{\sum_s R_s^2 s^2 n_s}{\sum_s s^2 n_s}, 
\end{equation}
where $n_s$ is the number of domains or clusters of size $s$ and the radius of 
gyration $R_s$ is defined as
$ R_s^2 = \sum_{i=1}^s \mid r_i - r_0 \mid^2/s $, 
where $r_i$ is the position vector of the $i$th spin of the cluster and
$ r_0 = \sum^s_{i=1} (r_i/s)$  
is defined as the centre of mass of the particular cluster. The 
pseudo-correlation length 
$\tilde \xi$ is observed to grow to system size order as one approaches the phase boundary (Fig. 3); thereby
providing further indication of the growth of a length scale. It should 
 be noted that, as in the 
static transition in the pure Ising system, the length $\tilde \xi$ is 
distinct from the correlation length \cite{robin}.


In the linear limit of the mean field dynamics, it has been possible to 
show the divergence of both the
length and time scales at the magnetization-reversal transition phase boundary. 
Sharp growth of these scales has also been observed in the Monte Carlo
case, studied in two dimension. Here, we looked at the size distribution
of the clusters or domains of reversed spins whose average size was observed
to grow at the phase boundary of the transition.


AM would like to thank A. Dutta for useful discussions. BKC is grateful to the 
INSA-Royal Society Exchange Program for supporting his visit to the Department
of Physics, University of Oxford, UK, where part of the work was done.



\section*{Figure Captions}

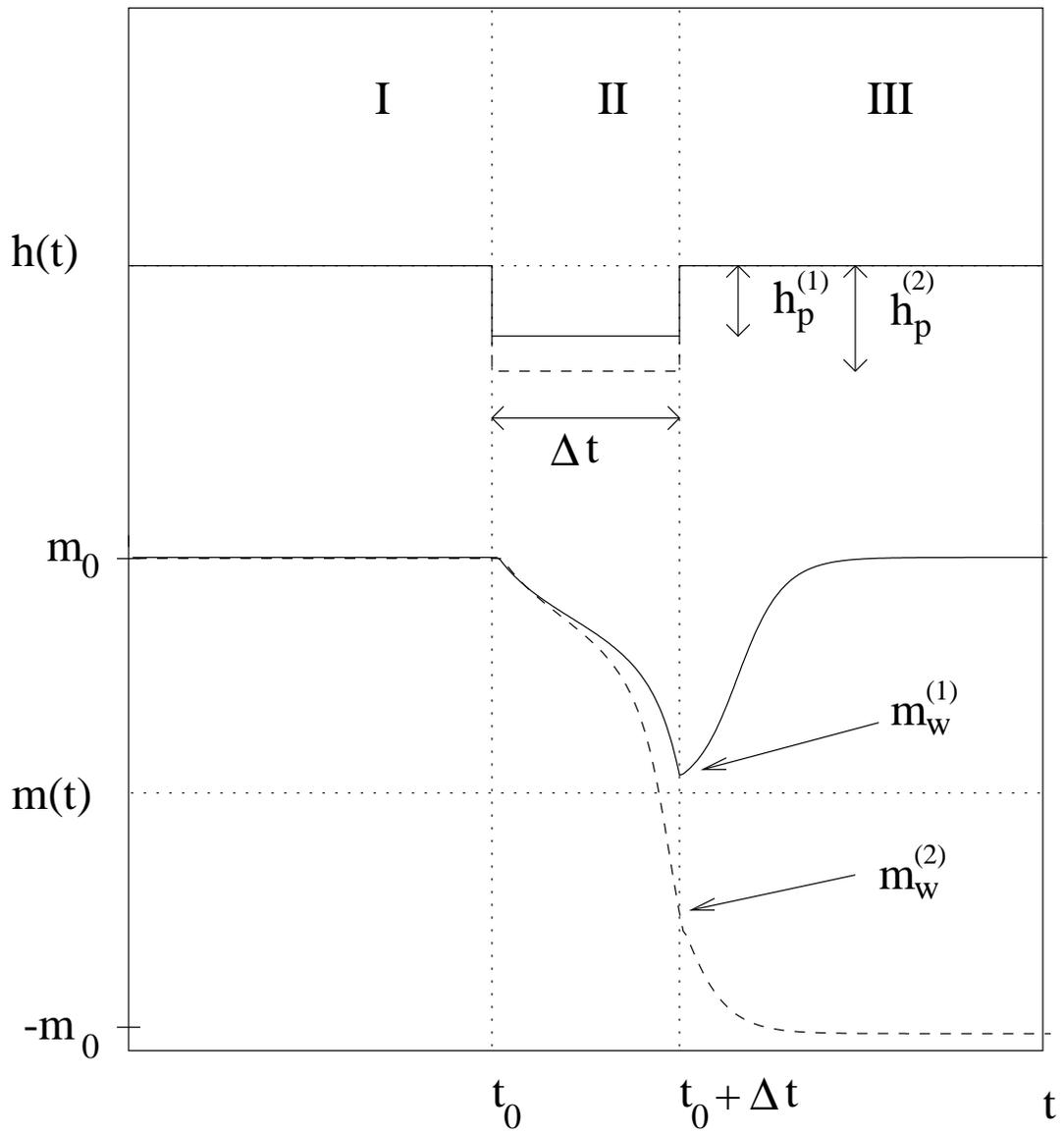
FIG. 1. Schematic time variation of the pulsed field $h(t)$ and the 
corresponding response magnetization $m(t)$ for two different cases.
The solid line indicates no magnetization-reversal case whereas the dashed 
line indicates a magnetization-reversal.

\vskip 1cm
\noindent
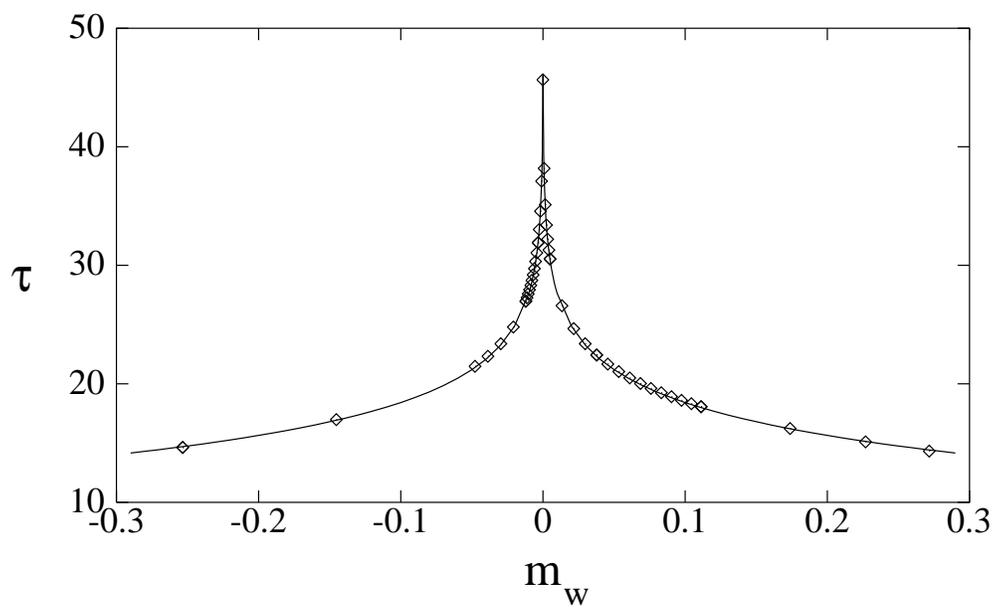
FIG. 2. Divergence of the relaxation time in mean field limit for $T=0.8$
and $\Delta t=20$ (from numerical solution of eqn. (4)). The solid line
indicates the corresponding analytical estimate (eqn. (15)).

\vskip 1cm
\noindent
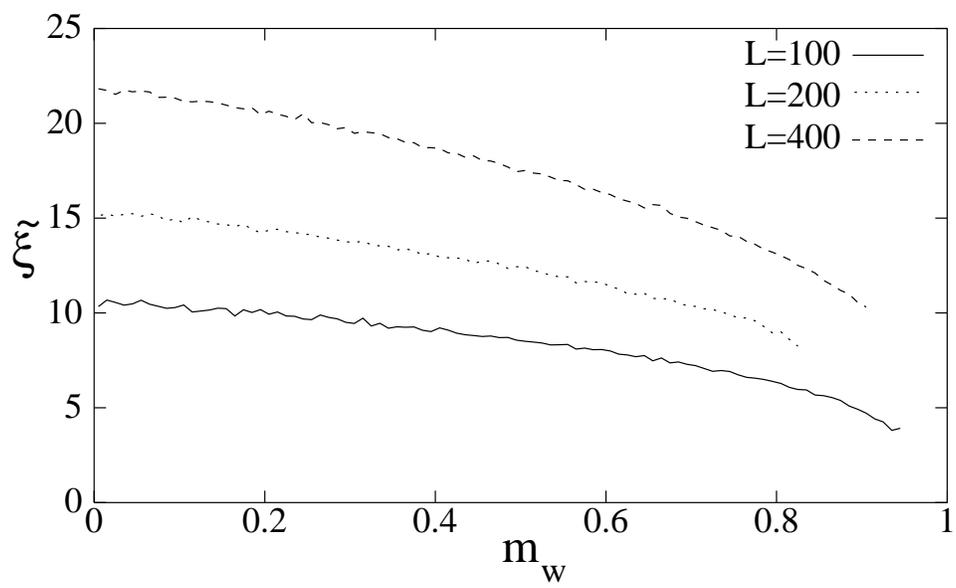
FIG. 3. Growth of the pseudo-correlation length $ \tilde \xi$ for different 
system sizes in the Monte Carlo study on a square lattice of size $L \times L$.
\end{document}